\documentclass[reprint,
notitlepage, amsmath,amssymb,
 aps,
final
]{revtex4-2}

\usepackage{mathtools}
\usepackage{graphicx,xcolor}
\usepackage{dcolumn}
\usepackage{bm}
\usepackage{hyperref}
\hypersetup{colorlinks=true,linkcolor=blue!90!gray,citecolor=blue!80!green!70!gray,filecolor=cyan!80!blue,urlcolor=teal!80!blue} 
\usepackage{nicefrac}
\usepackage[activate={true,nocompatibility},final,tracking=true,kerning=true,spacing=true,factor=1100,stretch=10,shrink=10]{microtype}

\usepackage{lmodern,booktabs,multirow,tabularx} \newcolumntype{Y}{>{\centering\arraybackslash}X}
\let\vec\bm

\newcommand{\Pe}{\mathrm{Pe}}

\renewcommand{\exp}[1]{\operatorname{e}^{#1}}
\newcommand{\ii}{\operatorname{i}}

\let\epsilon\varepsilon
\let\theta\vartheta
\usepackage{braket}

\begin{document}

\title{Spectral insights into active matter: Exceptional Points and the Mathieu equation}
 
\author{Horst-Holger Boltz}
\email[]{horst-holger.boltz@uni-greifswald.de}

\author{Thomas Ihle}

\affiliation{Institute for Physics, Greifswald University, 
17489 Greifswald, Germany}

\date{\today}

\begin{abstract}
    We show that recent numerical findings of universal scaling relations in systems of noisy, aligning self-propelled particles by Kürsten [Kürsten, arXiv:2402.18711v2 [cond-mat.soft] (2024)] can robustly be explained by perturbation theory and known results for the Mathieu equation with purely imaginary parameter. In particular, we highlight the significance of a cascade of exceptional points that leads to non-trivial fractional scaling exponents in the singular-perturbation limit of high activity. Crucially, these features are rooted in the Fokker-Planck operator corresponding to free self-propulsion. This can be viewed as a dynamical phase transition in the dynamics of noisy active matter. We also predict that these scaling relations depend on the symmetry of the alignment interactions and discuss the relevance of this structure in the free propagation for self-alignment and cohesion-type interactions.
\end{abstract}

\maketitle 

\nocite{kuerstennumerical}
\section{Introduction}
Models of active matter~\cite{ramaswamy2010,marchetti2010,menzel2012,bechinger2016,chate2020,essafri2022,tevrugt} have generally two aspects~\cite{vrugt2025} that robustly result in new physics as they lead to effective, generalized dynamics not beholden to the general structural framework of fundamental physical laws. They are open systems with an insertion of energy on the smallest scale that drives motion or generates forces, and they are coarse-grained theories that do not account for the internal mechanisms below a certain cutoff, usually this is done on the scale of active particles or agents whose internal operations are mostly outside the scope of modeling. These aspects manifest themselves in models with broken parity-time-symmetry ($\mathcal{P}\mathcal{T}$): there is a distinguished direction of time.

The more generalized structure of active matter models has immediate consequences for statistical physics approaches to the quantitative description of their collective behavior. There is almost always~\cite{morse,cameron,boltz2023,schüttler2025} no potential in a general sense (energy, thermodynamic potential, Lyapunov function) that  would allow for a direct formulation of the stationary measure in the vein of the Boltzmann weights of equilibrium thermodynamics. This has led to a derivation of adapted methodologies that are not reliant on assumptions of the model dynamics, such as first-principles kinetic theory~\cite{patelli2021,ihle2023,boltz2024}, effective mesoscale kinetic theory~\cite{bertin2006,bertin2009,peshkov2014,ito2025}, density functional theory~\cite{dean1996,archer2004dynamical,te2020classical,Vrugt_2023,illien2025} and symmetry-informed hydrodynamic equations~\cite{toner1998,toner2005,toner2024}. These approaches have complementary strengths and differ in the underlying assumptions and applicability ranges. One aspect of active matter models where this is particularly evident is the role of noise. As the dynamics of an active system are effective dynamics on a coarse-grained scale~\cite{vrugt2025}, they are in general manifestly stochastic even in a purely classical picture. This is due to the missing degrees of freedom that are lost in coarse-graining. Noise in active systems can be highly relevant to achieve functionality~\cite{zheng2022,chatterjee2025,liu2025}. While, for example, kinetic theory can be adapted~\cite{boltz2026} to finite noise strengths, it is conceptually a high-activity approach~\cite{pintogoldberg} that treats noise as a correction, other approaches, however, are operating in a  low-activity limit~\cite{loi2011,wang2011,maggi2015,fodor2016,wittmann2019} and treat the activity as a perturbation to an otherwise thermal system.

Details of the statistical theory approach taken aside, there is a common consequence of the more generalized dynamics in active matter systems that is intrinsically linked to the broken $\mathcal{P}\mathcal{T}$-symmetry: the emergence of  Non-Hermitian~\cite{trefethen,elganainy2018,ashida2020} operators. Non-Hermitian operators and matrices can, and in general do, show phenomena that are impossible in the usual Hermitian setting. One highly relevant example of this are exceptional points~\cite{kato2013perturbation,Heiss_2012}: upon change of a parameter the operator becomes defective, the geometric multiplicity of an eigenvalue is less than its algebraic multiplicity. In simple terms, this means that two (or more) of the eigenvalues collide in the complex plane and the associated eigenvectors are no longer orthogonal (possibly coinciding). A first point of contact with this concept can be found in simple mechanic systems~\cite{Dolfo_2018,Fernandez_2018} and many real-life applications exist in the realm of open or driven quantum systems, where particularly the enhanced sensitivity close to an exceptional point is of interest~\cite{wiersig}. From a theory perspective, exceptional points as a topological feature of the spectrum offer an opportunity for universal statements that are not subject to system details, as such there has been a surge of interest in recent years in active systems with exceptional points~\cite{sone2020,saha2020,fruchart2021,shankar2022,suchanek2023,suchanek2023a,suchanek2023b,fruchart2023,thiele2023,mecke2024,lama2025,kreienkamp2025,duan2025}.

A novel phenomenology of interest that motivated~\cite{reynolds,vicsek1995} a considerable amount of interest in active matter is flocking~\cite{toner2005,marchetti2010,vicsek2012,chate2020,toner2024}, the emergence of collective motion in ensemble of self-propelled agents. While such motion is readily observable in systems like flocks, swarms, herds or schools of fish, it is an emergent collective phenomenon conceptually intriguing. They are instances of system-wide orientational order in low dimensions ($d=2,3$), every agent on itself moves along an internally determined direction and these directions have to align for motion to emerge on the collective level. Truly long-ranged order in a continuous degree of freedom in two dimensions is only possible in active systems and can directly be traced back to the irreversibility~\cite{fodor2016,tasaki2020,dadhichi2020,ferretti2022} induced by activity, whereas such spontaneous breaking of a continuous symmetry in two dimensions is not possible in equilibrium systems~\cite{hohenberg1967,merminwagner1966}. The first evidence of such a transition was found in Vicsek models~\cite{ginelli2016,chate2020}, both numerically~\cite{vicsek1995} and in renormalization group analysis~\cite{toner1995,toner1998}. This is a class of paradigmatic toy models in which point particles are endowed with an internal orientation and the dynamics consist of streaming along this orientation with a fixed speed, aligning of the orientation to that of neighboring particles and orientational noise that is in competition with the alignment. As such the agents in Vicsek-like models are inherently polar, they have a distinct orientation. The ideas, concepts and methods developed in these simple toy models have since been adapted to study a large variety of systems such as migrating cells~\cite{mayor2016,Hakim_2017,alert_2020}, granular matter~\cite{ramaswamy2010,kumar_2014,bechinger2016,soni2020}, robotic systems~\cite{wiandt_2015,dorigo_2021,duan2023}, traffic~\cite{albi_2019} as well as other social systems~\cite{castellano2009} and, naturally, flocks of birds~\cite{cavagna_2014}. For this work, it is particularly relevant that a vast number of models share the same structure~\cite{chate2020} that was already present in the  original Vicsek model with respect to the self-propulsion, an internal orientation that is followed with (almost) fixed speed, and mostly differ in the interactions.

Remarkably it was found for certain birds that their aligning is best modeled if the criterion that determines if two agents consider each other neighbors is not the spatial distance (metric model), but the topological distance~\cite{ballerini2008,cavagna_2014,shankar2022}. This has inspired interest in metric-free models that consider interactions with topologically determined ranges, such as considering the $k$ nearest agents rather than agents within a pre-determined fixed range. From a theoretical perspective, these metric-free models are interesting as they seem to offer a way to study flocking with less significance of density fluctuations. Metric Vicsek-like models show a very pronounced coupling of the order and the density which leads to long-wavelength instabilities that are observable in form of, for example, anomalous density fluctuations or density waves~\cite{ginelli2016,zhao2021}. Many renormalization group studies are done in the unfortunately named class of Malthusian flocks, incompressible flocks with an unspecified but fast mechanism (death, birth) that equalizes any density fluctuations, to avoid the technical complications arising from density-order coupling~\cite{toner2024}. Whether the transition to order in metric-free models is continuous~\cite{ginelli2010,peshkov2012,chou2015} or discontinuous due to relevant fluctuations~\cite{rahmani2021,martin2021,Martin_2024} has been the subject of substantial seemingly incongruous debate. Recently, \citeauthor{zhao2025}~\cite{zhao2025} made important progress on this issue and found that there are indeed two different transitions. A continuous transition at low activity or high noise and a discontinuous transition at large activity or low noise. The relative strength of the activity is commonly expressed by means of the non-dimensional Peclét number $\mathrm{Pe}$ which we will introduce below. It fully describes the free propagation of self-propelled particles with rotational diffusion (active Brownian particles) as it is the only parameter in the non-dimensionalized free equations. The alignment interaction has an analogously non-dimensionalized typical strength which we call $\Gamma$. The discontinuous crictiality at large activity was found to obey a scaling relation. The critical coupling $\Gamma_c$, the onset of order, and the activity $\Pe$ seem to follow a simple power-law, using our notation this can be expressed as $\Gamma_c\sim \Pe^{\gamma}$ with a new exponent $\gamma \approx 2/3$. This criticality had been previously missed due to the way the parameter space was explored. The seemingly robust scaling relation is fairly remarkable. While scaling relations and critical exponents are ubiquitous in equilibrium critical phenomenology, such relations are rare in active systems and the origin of the simple rational exponent is not obvious. The low activity criticality is accessible via direct perturbation theory which has been noted before~\cite{escaff2020,escaff2024,escaff2025b}.

With this work, we want to elucidate some recent numerical results by \citeauthor{kuerstennumerical}~\cite{kuerstennumerical} from the perspective of exceptional points.  Kürsten sought to understand the observations by \citeauthor{zhao2025}~\cite{zhao2025} by means of kinetic theory. He performed a ring-kinetic closure of the metric-free model and numerically analyzed the spectrum of the operator for the two-particle function to find the origin of this relation. The transition from a disordered state to polar order can be understood as an instability of the slowest eigenmode that is excited by the interactions. The scaling is then derived as the result of numerically inferred scaling relations for the eigenvalue and projection to the polar mode of the eigenmode. We summarize Kürsten's findings in more detail below, but the instability approach correctly identifies the existence of two different scaling regimes for low and high activity and makes quantitatively correct predictions for the critical coupling strength based on the numerically inferred exponents.

We rephrase the results for the free two-particle dynamics as a consequence of the exceptional spectrum of the one-particle Fokker-Planck operator. Crucially, this analysis is applicable both for small and large activity. By mapping to a purely imaginary Mathieu equation, we can directly explain the phenomenological scaling behavior and give exact non-trivial values. The Mathieu equation arose originally in the context of elliptic sound drums~\cite{mathieu1868memoire}. In this form with real parameter it has many physical applications~\cite{ruby}, such as the vertically driven pendulum~\cite{dynamics}. Its solutions, the Mathieu functions, are subject to several monographs~\cite{mechel,mclachlan,arscott,meixner} as well as part of standard collections on special functions.~\cite{dlmf, wangguo} A gentle didactic introduction can be found in ref.~\cite{chaos2002}, while details of their numerical computation and historical context can be found in a recent review~\cite{brimacombe2021}. However, there has also been a long-standing~\cite{mulholland} interest in the Mathieu equation for complex or purely imaginary parameters. In particular, the existence of double points in the characteristic value~\cite{blanch1969,berezman,shivakumar} which correspond to exceptional points in the context used here has been studied a lot. A recent review of the Mathieu functions for imaginary parameter that contains the relevant results needed here can be found in Ref.~\cite{ziener2012}.

The relevance of the Mathieu equation for active~\cite{kurzthaler2018,mayer2021,reichert2021,zhang_2022,mason2023,zhang_2024,baouche_2024}  and $\mathcal{P}\mathcal{T}$-broken systems~\cite{bender2011,fernandez2014,fring_2015} has been recognized before, but to our knowledge the universal implications of the structure of these functions and their spectrum for aligning active matter have only been recognized in the numerical work by Kürsten~\cite{kuerstennumerical}.

The goal of this paper is not to only refer to the relevant literature results from the theory of the Mathieu equation, but to make these more accessible by also considering a finite-dimensional linear algebra approach. We present results from this approach that are intuitively accessible alongside with the respective literature results on Mathieu functions. As we consider properties of the free propagator of active models, these insights are relevant to the entire class of dry active dilute active matter~\cite{chate2020} and, particularly, universal with respect to the model specific interactions.

This paper is structured as follows. In the following section we will introduce a generic model of self-propelled particles in two spatial dimensions with alignment interactions and rotational noise. We rephrase this into a general form by rescaling. We will then briefly summarize the numerical findings of Kürsten~\cite{kuerstennumerical} to clearly motivate the analysis using the model specific language. In the next section, the spectrum of the free Fokker-Planck operator of this model is considered and the mapping to the Mathieu equation is introduced.  We will derive the relevant results by means of simple, accessible methods but also back this up by exact result from the theory of Mathieu functions. We close with a discussion of the implications of this work.

\section{Model}

We are considering models of self-propelled particles in two dimensions with fixed uniform speed $v_0$. Each particle $i$ ($i=1,\ldots,N$) has a position $\vec r_i$ and an internal orientation parameterized by an orientational angle $\phi_i$ with respect to an arbitrary reference axis. The dynamics we are interested in are given by
\begin{subequations} \label{eq:model}
\begin{align}
\partial_t \vec{r}_i &= v_0 (\cos\phi_i,\sin\phi_i)^{\mathrm{T}}\\
\partial_t \phi_i &= \sqrt{2 D_r} \xi_i + \Gamma \!\!\! \sum_{j\neq i, j\in \Omega_i} \!\!\! \mathcal{F}_{ij}(\vec{r}_i-\vec{r}_j,\phi_i,\phi_j,\lvert\Omega_i\rvert)
\end{align}
wherein the $\xi_i$ are white uncorrelated noise with 
\begin{align}
\langle \xi_i \rangle &=0\\
\langle \xi_i(t) \xi_j(t') \rangle &= \delta(t-t') \text{.}
\end{align}
\end{subequations}
An implicit control parameter for this system is the density which is given by the system size. We assume linear extensions $L$ in both dimensions with periodic boundary conditions, implying a number density of $\rho=N/L^2$. As written, we have considered a generic two-particle orientation interaction $\mathcal{F}_{ij}$ that can depend on pairwise distances and the involved angles. Examples that have been considered in the literature are alignment interactions of polar~\cite{peruani2008,ihle2023} or higher symmetry~\cite{nagai2015,boltz2024} ``turn-away''~\cite{das2024} or ``cohesion''~\cite{singh,shea2025} type torques that orient the particle with respect to the distance and vision-based alignment~\cite{lavergne2019}. These can have an explicit (metric) dependence on the pair distance or have more complex adjacency rules such as so-called metric-free or topological alignment models~\cite{cavagna2010,zhao2025} that consider alignment with the $k$ next neighbors. We allow for this by restricting the summation over interaction partners to some set of indices $\Omega_i$ that contains the adjacency information. The details of the interaction are not important to the discussion here and  we focus on the gestalt of the free propagation which is that of an active Brownian particle displaying self-propulsion and purely orientational diffusive noise. This is an aspect of these active models~\cite{chate2020} that is far less varied. The only notable exception are models that allow for variation in particle speed~\cite{mishra2012}. We will discuss this at the end.

For the time being, we study the free problem, i.e. $\Gamma \equiv 0$. In this case, we can identify $\tau_r=D_r^{-1}$ as the only relevant timescale of the free dynamics. This is the time over which the orientation of particles decorrelates by the rotational noise, i.e. propagation is persistent over stretches that are typically of duration $\tau_r$. Additionally, any interaction will introduce a typical lengthscale $\ell$. For metric interactions, this would be the particles' interaction radius, whereas for metric-free (or topological) interactions the typical distance $\ell^{-1}=\sqrt{\rho}=\sqrt{N}/L$ is the lengthscale of the interaction.   We rescale the units accordingly to non-dimensionalize the problem, such that $t \to  D_r t$ and $\vec r \to \vec r/\ell$. For the sake of notational clarity, we do not explicitly change the notation between the dimensional physical and rescaled dimensionless units. The free dynamics are then
\begin{subequations} \label{eq:peclet}
\begin{align}
\partial_t \vec{r}_i &= \Pe (\cos\phi_i,\sin\phi_i)^{\mathrm{T}}\\
\partial_t \phi_i &= \sqrt{2} \xi_i \text{.}
\end{align}
\end{subequations}

The Péclet number $\Pe = v_0 / (D_r \ell)$ is the one relevant quantity characterizing the single-particle behavior. The dynamics of eq.~\eqref{eq:peclet} has been labeled {\em active Brownian particles}~\cite{howse2007,hagen2011,romanczuk2012,basu2018}. The rotational diffusion of the orientational angle leads to a super-diffusive unordered motion of the agents. The important difference to standard Brownian motion is that this dynamic breaks $\mathcal{PT}$-symmetry, as an additional phase shift of $\pi$ in the angles $\phi_i$ is needed to actually invert the dynamics.

\section{Previous numerical results}

To remain self-contained, we give a brief summary of the numerical findings~\cite{kuerstennumerical} that motivated this work. These are scaling relations for the spectrum of the free two-particle propagator, below we reduce this to the one-particle Fokker-Planck operator. The idea of Kürsten is that the onset of order can be understood as an {\em instability of the relaxational dynamics} under excitation by the interactions. Thus, the key is insight into the relaxational dynamics around the unordered state that is the stationary solution of the free problem. To this end, there is a particular interest in the eigenmode of the free propagator that features the slowest relaxation, i.e. has the eigenvalue with the largest real-part. In the free problem, all eigenvalues will be non-positive. Calling this eigenvalue $\lambda_0$, it was found numerically that
\begin{align}
\operatorname{Re}(\lambda_0) \sim \begin{cases} \Pe^{\alpha_\text{low}} & \text{for } \Pe\ll 1 \\
   \Pe^{\alpha_\text{high}} & \text{for } \Pe\gg 1\end{cases} \label{eq:lambda}
\end{align} 
with scaling exponents $\alpha_\text{low} \approx 2$ and $\alpha_\text{high}\approx 0.5$ characterizing the limits of asymptotically small and large activity, respectively.

Many alignment interactions have polar symmetry. Therefore, Kürsten also considered the projection of the eigenmode that corresponds to $\lambda_0$ onto the polar angular mode. This measures to what extent the long-living mode in this system is excited by interactions. Calling this projection, which we formalize later on, $c_0$, the numerical results are
\begin{align}
c_0 \sim \begin{cases} \Pe^{\beta_\text{low}} & \text{for } \Pe\ll 1 \\
   \Pe^{\beta_\text{high}} & \text{for } \Pe\gg 1\end{cases} \label{eq:proj}
\end{align}
with exponents $\beta_\text{low} \approx 1$ and $\beta_\text{high}\approx -0.125$. Rather intriguingly, the numerical results are all simple rational values within numerical precision. We can corroborate this using our own numerics, see figs.~\ref{fig:projection} and \ref{fig:spectrum}. This suggests a robust general mechanism behind these values, which indeed is what we find. Importantly, the exponents $\alpha, \beta$ can be used to accurately predict scaling relations for the critical coupling strength in interacting systems and these are in agreement with direct agent-based simulations~\cite{zhao2025}. In the following section, we will show that the given values are indeed exact and can be understood by means of perturbation theory and exceptional points.

\section{Spectrum of the free Fokker-Planck operator}

The stochastic dynamics of eq.~\eqref{eq:peclet} are of Langevin type and can therefore be equivalently rephrased as a Fokker-Planck-equation for the $N$-particle distribution $P_N$. Since we are considering free, non-interacting particles, we know that the particles are independent, molecular chaos holds, and the $N$-particle distribution trivially factorizes $P_N(\{\vec r_i, \theta_i\},t)=\prod_i f(\vec r_i,\theta_i,t)$ with $f$ being the one-particle function. Its time evolution is given by the one-particle Fokker-Planck equation (or the first member of the BBGKY ``hierarchy'' that is fully decoupled in the free problem)
\begin{align}\label{eq:fp3}\begin{split}
&\partial_t f(\vec r, \phi,t)=\mathcal{L} f(\vec r,\phi) \\&= \left(- \Pe(\partial_x \cos\phi + \partial_y \sin\phi) + \partial_\phi^2 \right) f(\vec r,\phi,t) \text{.}\end{split}
\end{align}
In light of the inherent symmetry under spatial translations, it is more natural to consider the dynamics after Fourier transforming the positions $f(\vec r, \phi,t)=\int\! \frac{\mathrm{d}\vec k}{(2\pi)^2} \exp{-\ii \vec r \cdot \vec k} f(\vec k,\phi,t)$. The Fourier modes trivially decouple for symmetry reasons and we can choose without any loss of generality $\vec k =(k,0)$ to find
\begin{align}
\mathcal{L} f(k,\phi,t) &= \left(\ii \Pe\, k \cos\phi + \partial_\phi^2 \right) f(\vec k,\phi, t) \text{.} \label{eq:fp1}
\end{align}

In particular and following the logic of Kürsten~\cite{kuerstennumerical} albeit slightly more direct, we are interested in the temporal eigenmodes, that is we are looking for a solution to the eigenvalue equation
\begin{align}
\partial_t f(k,\phi,t) &= - \lambda f(k,\phi,t)
\end{align}    
which allows us to directly eliminate the time dependence from eq.~\eqref{eq:fp1}.  The equation of interest therefore is
\begin{align}
0 &= \left(\lambda + \ii \Pe\, k \cos\phi + \partial_\phi^2 \right) f(\phi) \text{.} \label{eq:fp1_2}
\end{align}
We have also dropped the dependence on the wave number $k$ from the one-particle function as it only occurs as an effective parameter. In fact, this equation eq.~\eqref{eq:fp1_2} is rather well known already as it is essentially the Mathieu equation with purely imaginary parameter. To make this clearer, we substitute $\phi=2\psi$ and introduce the new notation $a=4\lambda$ and $q=2 k \Pe$ and find
\begin{align}
0 &= (a + \ii 2 q \cos 2\psi + \partial_\psi^2) f(\psi) \label{eq:mathieu} \text{.}
\end{align}
This is the Mathieu equation~\cite{dlmf,ziener2012}. We note that we treat here $q$ as the real parameter of the complex Mathieu equation, while other works consider $-\ii q$ as the imaginary parameter of the general Mathieu equation. This can lead to slight deviations in notation. It is a special case of Hill's differential equation~\cite{hill1877,magnus2013,dlmf}, $\ddot{y}+f(t)y=0$ with $\pi$-periodic $f(t)=f(t+\pi)$. As the coefficients of this ordinary differential equation are periodic, substantial insight can be drawn from Floquet theory, in particular by virtue of Floquet's and Ince's theorems~\cite{arscott}. In brief, there are (anti-)symmetric solutions (also called basically periodic solutions) with $f(\psi)=\pm f(\psi+\pi)$. For special values $a(q)$, that are called the {\em characteristic values}, there is one periodic solution and by virtue of the equation being second-order another non-periodic solution that is not of interest to us here. These periodic solutions are the {\em Mathieu functions of the first kind}:  the cosine-elliptic  and sine-elliptic functions which are typically denoted by $\operatorname{ce}_n$ and $\operatorname{se}_n$ with discrete index $n=0,1,2\ldots$, respectively. The name correlates with the notion that for sufficiently small $q$, these solutions are similar to sine and cosine or higher harmonics thereof. We are interested in even solutions with $\pi$-periodicity, $f(\psi)=f(\psi+\pi)$, as only these are admissible for the original problem, this means~\cite{dlmf} that the solutions of interest are the cosine-elliptic functions to even number $\operatorname{ce}_{2n}$. We will introduce further results of the theory of Mathieu functions as needed below, but refer to the literature for details outside of the scope of this presentation~\cite{mechel,mclachlan,arscott,meixner,brimacombe2021,ziener2012} . The Mathieu parameter $q=2k \Pe$ depends on the wave length $k$ and the Péclet number. We postpone a discussion of the wave-length dependence and, for now, treat $q$ as a measure of activity with $q=0$ being a diffusive system and $q\to \infty$ being the deterministic limit. 

While we can build on established results for the Mathieu equation, it is also instructive to consider a complementary route that allows to find some results directly with rather basic ``pedestrian'', linear-algebra tools. If we also Fourier transform with respect to the internal orientation, $f(\psi) = \sum_m \exp{\ii m \psi} f_m$, the problem reduces to an infinitely dimensional eigenvalue problem~\cite{escaff2020,kuerstennumerical,escaff2024,escaff2025,escaff2025b}
\begin{align}
0 &= (a-m^2)f_m + \frac{\ii q}{2} (f_{m-2}+f_{m+2})  \label{eq:eigenvaluematrix}
\end{align}
or
\begin{align}
a \vec f &= \mathsf{M} \vec f
\end{align}
with $(\vec f)_m=f_m$. We can functionally ignore the modes with odd index $m$ as these do not have the correct $\psi \to \psi+\pi$ symmetry (as the physical angle $\phi$ is twice the Mathieu angle $\psi$). In this rephrasing, it is perhaps the most obvious that as a consequence of the $\mathcal{PT}$-symmetry-breaking the occurring operators are Non-Hermitian. The matrix $\mathsf{M}$ is symmetric and tridiagonal (if the subspace of even modes is considered), but the off-diagonal elements are purely imaginary and, thus, change sign under conjugation. As a consequence, the usual constraints on the spectrum that follow from dealing with Hermitian matrices do not hold: The eigenvalues $\lambda$ can and generally will be complex and multiple eigenvalues can cross upon change of a parameter. In particular, exceptional points can occur, where multiple eigenvalues are identical with a single associated eigenvector~\cite{kato2013perturbation,Heiss_2012}.

From the structure of the matrix $\mathsf{M}$, we see that in the case of vanishing activity or asymptotically large noise, i.e. $\Pe =0$ and $q=0$, the eigenmodes are identical to the Fourier modes and the corresponding eigenvalues are $-m^2$. In the following, we consider the two asymptotic cases of small and large parameter $q$ in finite-dimensional approximations that is limited to $m=0,2,4,\ldots, 2Q$ with fixed integer cut-off $Q$ to achieve some direct insight. To alleviate the notation, we limit the analysis to the physical relevant case of even (w.r.t to $\psi$) and real functions. We denote this by transforming $f_m \to f_{m'} = (f_{2m}+f_{-2m})/2$. Then, the eigenvalue equation eq.~\eqref{eq:eigenvaluematrix} is
\begin{align}
0 &= (\lambda-m'^2) f_{m'} + \frac{\ii q}{2} (f_{m'-1}+f_{m'+1})   \label{eq:eigenvaluematrix2}
\end{align}
for $m'=0,1,2,\ldots,Q$.  Studying the linear algebra variant for finite $Q$ is additionally motivated by noting that $Q\approx N/2$ is enforced due to aliasing in a system with fixed particle number $N$, akin to the Nyquist criterion.

\subsection{Small activity, $q\ll 1$}

We approach the limit of small activity by considering the activity as a perturbation to the purely diffusive problem. Inspecting eq.~\eqref{eq:eigenvaluematrix2} for $q=0$, it is obvious that the Fourier modes are already eigenfunctions for the unperturbed problem and, therefore, we find the eigenvalues $\lambda_{m'}^{(0)}={m'}^2$. This holds for both the original modes as well as the combinations considered here. We can assess corrections to that  for small $q$ by means of standard perturbation theory.  We see directly, that the correction $\mathsf{M}'$ with $\mathsf{M}=\mathsf{M}^0+{\ii q\,}\mathsf{M}'$ only contains off-diagonal elements, that is $\bra{m'}\mathsf{M}'\ket{m'}=0$. Thus, there is no correction to first order. We note that this perturbative limit has been considered before analytically~\cite{escaff2020,escaff2024,escaff2025,escaff2025b}.

In second order perturbation theory and employing bra-ket notation (keeping in mind that these states are not constructed to be $L^2$-normalized) with $\ket{m'}=f_{m'}$, the eigenvalues change as
\begin{align}
      \begin{split}
\lambda_{m'}^{(2)}-\lambda_{m'}^{(0)}&= q^2 \!\sum_{n'\neq m'}\! \frac{\lvert \bra{n'}\mathsf{M}'\ket{m'}\rvert^2}{\lambda_{m'}^{(0)}-\lambda_{n'}^{(0)}} \\&=  q^2 \frac{1}{2(4m'^2-1)} \sim q^2 \text{.} \end{split}\label{eq:smallpperturb}
\end{align}
Crucially, the perturbation is non-degenerate within the subspace of modes considered here, whence this is indeed the leading order. Note that this scaling holds for any mode. This is not only consistent with the numerical findings of Kürsten~\cite{kuerstennumerical} rigorously establishing $\alpha_\text{low}=2$, but also holds in the limit $Q\to \infty$ as is evident from comparison with the characteristics of the Mathieu equation \eqref{eq:mathieu} in the limit of small $q$:~\cite{dlmf,frenkelportugal} 
\begin{align}
a_{2r} &= 4r^2 - \frac{q^2}{2(4r^2-1)} +\mathcal{O}(q^4) \text{.}
\end{align}
This result holds for even indices $2r$, corresponding to the even Floquet modes considered here.

The corresponding eigenmodes change accordingly to
\begin{align}
\ket{m'^{(1)}} &= \ket{m'} + q \sum_{n'\neq m'}\ket{n'} \frac{\bra{n'}\mathsf{M}'\ket{m'}}{m'^2-n'^2} \\ &= \ket{m'} + \ii \frac{q}{2} \left( \frac{\ket{m'-1}}{1+2m'} + \frac{\ket{m'+1}}{1-2m'} \right) \text{.}
\end{align}
As usual, the first order correction in the eigenstates resembles the second order correction to the eigenvalue. Kürsten~\cite{kuerstennumerical} considered in particular the projection of the lowest eigenstate to the polar mode. The lowest eigenmode is for sufficiently small values of $q$ (see below) the mode $\ket{0'}$ and, therefore, there is a new ``density-order-like'' coupling
\begin{align}
\braket{1'|0'^{(1)}} &= \ii q \frac{1}{2} \sim q \text{.}
\end{align}
The linear scaling, i.e. $\beta_{\text{low}}=1$, again agrees with the numerical results of Kürsten~\cite{kuerstennumerical}, but having access to the full analytical representation allows for further insight. Especially, we see that the scaling of this projection is not fully universal: If we instead considered the projection of the nematic mode (as is appropriate for interactions of nematic symmetry), this first order correction would vanish and the leading order would be the second order. In this way, the most frequently considered case of polar interactions are special in this low-activity limit as they connect to the polar symmetry of the propagation  and the generic case is different. We consider this in more detail when considering interacting systems below.

 \subsection{High activity, $q\gg Q^2$}
 We turn to the opposite asymptotical case of high activity. Trying to approach this analogously with the diffusion now as a perturbation to the active advection, we have to require that $q$ is large against the largest eigenvalue of the perturbation,  that is $q\gg Q^2$, for the noise to actually be a small perturbation. It is at this stage already obvious, that this will be critically ill-posed in the limit of $Q\to \infty$, but we postpone this discussion to later. This is a direct manifestation of the fact that the diffusivity (or inverse Péclet number) as a small parameter enters the Fokker-Planck equation as the coefficient of the highest (second-order) derivative and this is, therefore, a singular perturbation. Such a singular perturbation will lead to a boundary-layer phenomenology in the solutions, which will become apparent in the explicit expressions later on.

For $q^{-1}=0$, we are left with a tridiagonal Toeplitz matrix, which is defined by 
\begin{align}
\tilde{\mathcal{L}} \ket{m'} &= \ii (\ket{(m-1)'}+\ket{(m+1)'})
\end{align}
for $m'=1, \ldots, Q$ and $\tilde{\mathcal{L}} \ket{0'} = \ii \ket{1'}$. We can make use of the well-known~\cite{noschese} closed expression for the eigenvalues and -vectors of tridiagonal Toeplitz matrices. For $q^{-1}=0$, we find that the eigenmodes $\ket{n_q}$ and corresponding eigenvalues $\lambda_{q,n}$ are given by
\begin{align}
\ket{n_q} &\propto \sum_{m'} \sin\left( \frac{(m'+1)\,(n+1)\,\pi}{Q{+}2} \right) \ket{m'}\label{eq:toeplitzvectors}\\
\tilde{\lambda}_n^{(0)} &= -2 \ii \cos{\frac{(n+1)\pi}{Q{+}2}} \text{.}
\end{align}
As we can see, these are purely imaginary eigenvalues. So, the real part at any finite $q^{-1}$ is a result of perturbative corrections. The symmetry considerations of before do not apply and the first order correction is the leading term. This is directly computed to 
\begin{align}
    \begin{split}
&\mathrm{Re}(\tilde{\lambda}_n^{(1)})=q^{-1}(24 (Q{+}2))^{-1} \cdot \\&\cdot \left[-2 Q \left(2 Q^2+3 Q+1\right)+ \csc \frac{\pi }{Q{+}2}  \cdot \right.\\
&\left\{\left(6 Q-9 \cos \frac{4 \pi }{Q{+}2}+9\right) \csc\frac{\pi }{Q{+}2}\right.\\&{+}3 \left(\sin \frac{\pi  (1-2 Q)}{Q{+}2}\sin \frac{\pi }{Q{+}2}\right) \csc ^2\frac{\pi }{Q{+}2}\\&{+}\left.\left.6 \left(\sin \frac{\pi  (1{-}Q)}{Q{+}2}{-}(Q{+}1)^2 \sin \frac{\pi  (Q{+}1)}{Q{+}2}\right)\right\}\right]\end{split}
\end{align}
The relevant eigenvalue is still found for $n=0$ and takes some $Q$ dependent value. The leading order projection of this eigenstate to the polar mode will be a constant as is evident from eq.~\eqref{eq:toeplitzvectors}.

The results in this section are only on first glance in conflict with the numerical results of Kürsten~\cite{kuerstennumerical} as the perturbation approach only holds for $q \gg Q^2$ which is irrelevant in the continuum limit $Q\to \infty$. In this limit, any amount of noise constitutes a singular perturbation. However, the numerical observations~\cite{kuerstennumerical} were performed at finite $Q$ for large, but not asymptotically large values of $q$. So, even at finite $Q$ there has to be an in-between scaling regime that we refer to as the intermediate regime. We could directly refer to results for the Mathieu differential equation which correspond to the limit of infinite $Q$ and which we will give at the end of the discussion, but consider a more direct approach that aims to understand the nature of this regime intuitively to be more instructive.

\subsection{Intermediate regime, $1\ll q \ll Q^2$}

\begin{figure}
\includegraphics[width=\linewidth]{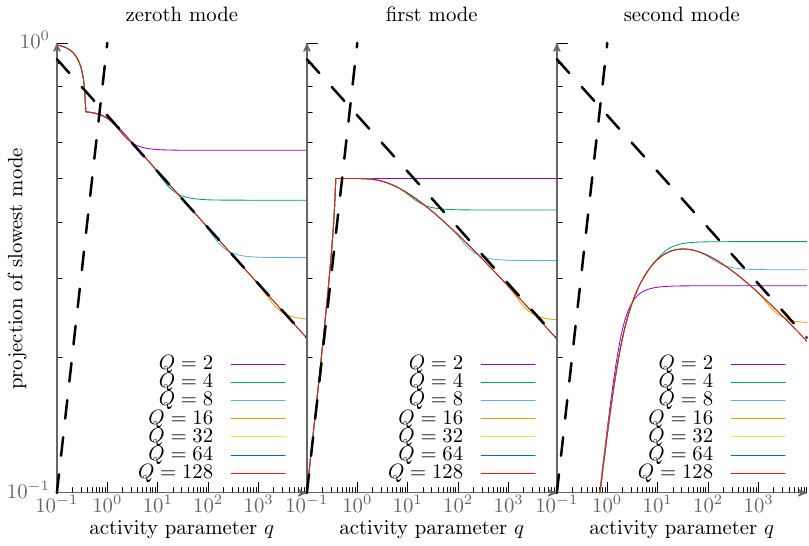}
\caption{Projection of the numerically found lowest eigenmode onto the lowest three Fourier modes as a function of the activity parameter $q$ and for various numbers $Q$ of considered modes, see text for detail. The axes use double-logarithmic scaling. We are also providing the scaling behaviors $q^1$ and $q^{-\frac{1}{8}}$ (black dashed lines) found in earlier numerical works~\cite{kuerstennumerical} as guides to the eye. The diagonalization of the system eq.~\eqref{eq:eigenvaluematrix} was done using the \textsc{ComplexEigenSolver} routines of the \textsc{Eigen} library~\cite{eigenweb}.}
\label{fig:projection}
\end{figure}

\begin{figure}
\includegraphics[width=\linewidth]{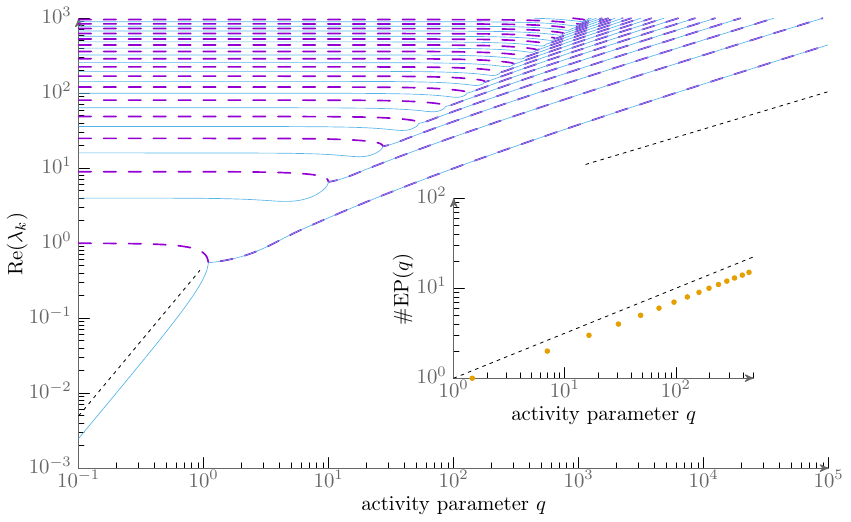}
\caption{Large: Real part of numerically determined eigenvalues for $Q=200$ shown in log-log format as a function of the control parameter $q$. The power-law relations of eq.~\eqref{eq:lambda} are indicated by dashed black lines. For visual clarity, we display every other eigenvalue differently, highlighting the coalescence of pairs of eigenvalues in a cascade of exceptional points. Inset: Number of EPs seen up to $q$. The number grows as $q^{\frac{1}{2}}$ which we indicate via a dashed black line. The diagonalization of the system eq.~\eqref{eq:eigenvaluematrix} underlying both figures was done using the \textsc{ComplexEigenSolver} routines of the \textsc{Eigen} library~\cite{eigenweb}.} \label{fig:spectrum}
\end{figure}

We present results for the projection from a direct numerical diagonalization in Fig.~\ref{fig:projection} of the slowest eigenmode to the first Fourier modes for various numbers of angular modes considered $Q$. From these it is obvservable, that there is indeed an intermediate regime in $q$ in which the non-trivial scaling $q^{-\frac{1}{8}}$ found by Kürsten is evident. A potential resolution to this is directly implied by the $\mathrm{Re}(\lambda) \sim q^{\approx 0.5}$ scaling found numerically. This scaling is a hallmark of a second order exceptional point~\cite{kato2013perturbation,Heiss_2012}, i.e. a point where the change in a control parameter of a Non-Hermitian matrix leads to two eigenvalues coalescing with a single eigenvector. This is ultimately equivalent to the typical scaling in pitchfork bifurcations (fold catastrophes). The existence of exceptional points in the purely imaginary Mathieu equation, double points of the characteristics, is long established~\cite{blanch1969,ziener2012}. We show the negative lowest few real parts of the eigenvalues as found from direct numerical diagonalization Fig.~\ref{fig:spectrum}. One observes a cascade of exceptional points each of which consists of two originally (for small $q$) distinct and real values meeting and having identical real parts after the exceptional point (but conjugated imaginary parts (not shown), that grow linearly in the distance). The usual expectation of the eigenvalue gap scaling with the root of the distance to the exceptional point thus applies to the {\em left} of the exceptional points or for smaller activity, but we (as Kürsten~\cite{kuerstennumerical}) find similar scaling also to the right for larger activity. However, this is not for the gap of the eigenvalues whose real part coincide but for their plain value and it is not close to the exceptional point but far away from it. Preempting the discussion below, we summarize the result as a guide for the following. The robust scaling seen far from the exceptional points at intermediate activity $1\ll q \ll Q^2$ is the result of coupling of many exceptional points. It is therefore a consequence of the full spectrum and cannot be understood by a local discussion in two modes. This global mechanism is rather special and not usually discussed as a consequence of the presence of exceptional points in the nonequilibrium physics literature.

 Already from the perturbative result in eq.~\eqref{eq:smallpperturb}, one sees that the eigenvalue of the zeroth mode is raised, while the eigenvalue of the first mode is lowered, meaning that they are seemingly poised to eventually meet. This is actually true and can be understood better in a low dimensional subspace. At some activity $q^*$ the two lowest eigenvalues collide and generate a non-vanishing imaginary part. Additionally, there is only a {\em single} eigenmode, this is only possible due to the Non-Hermitian structure. A very accessible version is the case of $Q=1$ which acts as a model exceptional point
\begin{align}
\mathsf{M}_{Q=1}&=\begin{pmatrix} 0 & \ii q/2 \\ \ii q/2 & -1 \end{pmatrix} \text{.}
\end{align} 
In this case, the eigenvalues $\lambda_{0,1}=\frac{1}{2}\pm \frac{1}{2}\sqrt{1-q^2}$ coalesc for $q=q^*= 1$. In this two-dimensional system, the exceptional mode is $\frac{1}{\sqrt{2}} \left( \ket{0} +\ii \ket{1} \right)$. Numerically, we find that the intermediate regime increases in size with taking into account more modes, see fig.~\ref{fig:projection}. Additionally, the critical activity converges very quickly. The value in the limit $Q\to \infty$ can be found by a numerical root search in the Mathieu characteristics $a_0(q)=a_2(q)$. We have done this using the implementations in the computer algebra system \textsc{Mathematica}~\cite{Mathematica} to find $q_c \approx 1.46876$ agreeing with literature values and our numerical results. As far as we aware, there is no closed expression for this value.

 As the activity is raised beyond this first exceptional point, the perturbation from the lower states and interaction with a cascade of exceptional points, the sequence of collisions of the eigenvalues $\lambda_{(2r)'}$ and $\lambda_{(2r+1)'}$, becomes relevant. This leads to a real part that moves like $\lambda \approx \lambda^* + c \sqrt{q-q^*} \sim \sqrt{q}$. Assessing the scaling in the intermediate regime which is the large activity case for the continuum case analytically is  highly non-trivial. The way of inferring the power-laws analytically is by means of a Newton-Puiseux series approach~\cite{kato2013perturbation} that allows for non-integer rational exponents in the control parameter. We note that the exceptional point density $\rho_q=\partial_q N \sim q^{-1/2}$ with $N(q)$ being the number of exceptional points below activity $q$ is already indicative that non-trivial power-laws can be relevant here, see inset of fig~\ref{fig:spectrum}.  In the spirit of accessibility of this note, we still try to point out the mechanism behind this scaling in the following subsubsection by considering a tailored low-dimensional model. Afterwards, we give results from mathematical Poincaré-type asymptotic analysis.
 
\subsubsection{Insight from model of coupled exceptional points}

\begin{figure}
    \includegraphics[width=\linewidth]{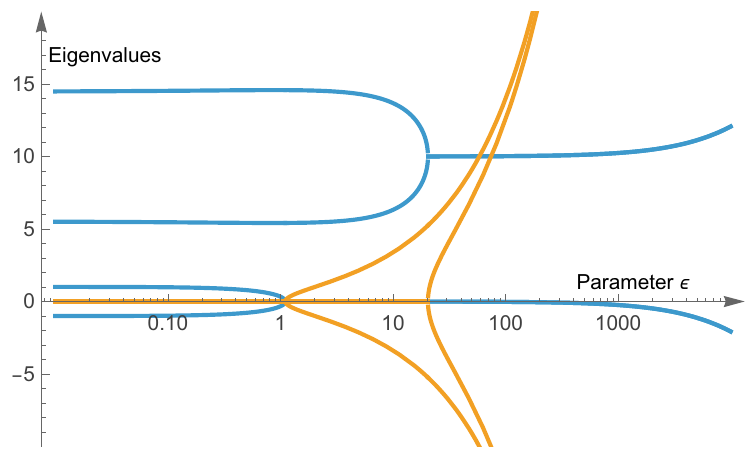}\\
    \hspace*{-0.3cm}\includegraphics[width=\linewidth]{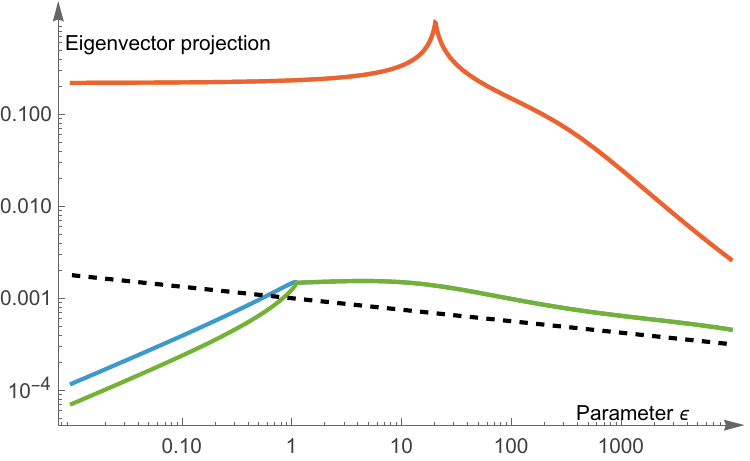}
    \caption{Numerical data for the system of coupled exceptional points in eq.~\eqref{eq:simple} obtained via direct diagonalization using the computer algebra system \textsc{Mathematica}~\cite{Mathematica}. Here, we use $c=0.1$. Top: Eigenvalues, the real part is shown in blue and the imaginary part in orange. The spectrum clearly has two exceptional points (slightly shifted due to the perturbative coupling). Bottom: Numerical evidence for $\varepsilon^{-\frac{1}{8}}$ scaling (as indicated by the dashed black line) in the projection of the eigenvectors to the $(0,0,0,1)$ mode. The different vectors are represented by differing colors with two curves coinciding (red line, corresponding to the upper two eigenvalues).}
    \label{fig:1over8simple}
\end{figure}

To see that the projection property $c_0=\braket{1|\vec{v}}\sim q^{-1/8}$ found numerically, is indeed attributable to the cascade of exceptional points without being reliant on deeper techniques such as Puiseux series, we consider the following toy system
\begin{align} \mathsf{M}(\varepsilon) &=
\begin{pmatrix} 
0 & 1-\varepsilon & 0 & 0\\
1 & 0 & 0 &0\\
0 &0 & 10 & 20 -\varepsilon\\
0 & 0 & 1 & 10
\end{pmatrix}
+ c \sqrt{\varepsilon} \begin{pmatrix}
0 & 1 & 0 & 0\\
1 & 0 & 1 & 0\\
0 & 1 & 0 & 1\\
0 & 0 & 1 & 0
\end{pmatrix} \text{.} \label{eq:simple} 
\end{align}
The idea behind this constructed matrix is that this system has originally two exceptional points at $\epsilon_1=1$ and $\epsilon_2=20$. These two subsystems are initially uncoupled and distanced by a rather large distance of $\Delta=10$. We emulate the effect of all the other exceptional points in the Mathieu system by introducing an effective coupling of the two exceptional points that scales with $\sqrt{\varepsilon}$. The scaling is reflective of the number of exceptional points found earlier, see inset of fig.~\ref{fig:spectrum}. While this is a simple system, it already can barely be solved analytically. We show numerical results in Fig.~\ref{fig:1over8simple}.

We find that an $\varepsilon^{-\frac{1}{8}}$ scaling in the projection of the lowest mode onto the unperturbed second mode is generated by the $\sqrt{\varepsilon}$ coupling of the modes. In this simple system it was inserted by hand, whereas in the full system it is a consequence of the eigenvalue scaling and the density of exceptional points. We note that the  perturbative behavior for low parameter values is not correctly modeled in this low-dimensional approach. This could be achieved by more detailed modeling choices. Here, however we only wanted to give a physical reasoning behind the emergence of non-trivial scaling laws that are due to {\em coupling of several} exceptional points, rather than the more standard power-laws found close to the exceptional point. We do not consider or intend eq.~\ref{eq:simple} to be a direct model for the system of eq.~\eqref{eq:eigenvaluematrix2}, the purpose of this model is to highlight the emergence of non-trivial, but rational scaling laws in the presence of multiple coupled exceptional points.

\subsubsection{Insight from Mathieu equation theory}

While the toy systems considered so far are insightful, their implications are far from conclusive. We therefore now will present the relevant literature results that allow for a robust analytical foundation of the exponents. For large parameters, the Poincaré-type asymptotics of the characteristic values and the solutions of the Mathieu equation are well established~\cite{dlmf,frenkelportugal}. Using the short-hand $s=2m+1$, one finds that the characteristic value is given by
\begin{align}
    \begin{split}
a_m(q)=&2 \ii q-(1+\ii) s \sqrt{2q}+\frac{s^2+1}{8} + \mathcal{O}(q^{-\frac{1}{2}})\end{split}
\end{align}
and, therefore, we directly see that
\begin{align}
\operatorname{Re(a_m)} &\sim \sqrt{q} \text{,}
\end{align}
i.e. $\alpha_\text{high}=\frac{1}{2}$. We want to stress again, that while this exponent is related to the presence of an exceptional point it is not due to closeness to the exceptional points, but rather the scaling {\em far away} from the relevant exceptional points due to the coupling to a cascade of exceptional points.

For the eigenmodes, we find  in the literature~\cite{dlmf} with the short-hand $\xi=2 (\ii q)^{\frac{1}{4}} \cos x$ with $x=\frac{\pi}{2} +\delta (\ii q)^{-\frac{1}{8}}$ wherein $\delta$ is a constant with $\lvert \delta\rvert < 2^{\frac{1}{4}}$
\begin{align}
\operatorname{ce}_{m}\left(x,q\right)=&\widehat{C}_{m}\left(U_{m}(\xi)+V_{m%
}(\xi)\right) \\
U_{m}(\xi)\sim& D_{m}\left(\xi\right)+\mathcal{O}(q^{-\frac{1}{2}})\\
V_{m}(\xi)\sim& \mathcal{O}(q^{-\frac{1}{2}})
\\
\widehat{C}_{m}\sim&\left(\frac{\pi \sqrt{\ii q}}{2(m!)^{2}}\right)^{\frac{1}{4}}\left(\tilde{C}_m\right)^{-\frac{1}{2}}\\
\tilde{C}_m=& 1+\frac{2m+1}{8\sqrt{\ii q}}+\mathcal{O}(q^{-1})
\end{align}
Herein, $D_m$ are the parabolic cylinder functions~\cite{dlmf}. While we omit details here, there are two important things to notice. For one, the typical width of the eigenmode (as indicated by the scaling in the definition of the $x$ variable that is reflective of a boundary layer) scales as $q^{-\frac{1}{8}}$ and, for two, the constant $\widehat{C}_{m}\sim q^{-\frac{1}{8}}$ are the normalization (cp. eq. (58) in ref.~\cite{frenkelportugal}). Thus, even on this very rough level, it seems plausible that the relevant scaling of the Fourier modes is indeed also of the form $q^{-\frac{1}{8}}$.  However, there are explicit results for the Fourier components of the solutions to the purely imaginary Mathieu equation~\cite{wolf2008asymptotic,ziener2012} that we can employ. Specifically, it was found that in our notation the following expressions are asymptotically correct for large $q\gg 1$
\begin{align}
    \braket{(2l)' | 0'(q)} &\approx (-1)^l \frac{\sqrt{(2l)!}}{l! 4^{\frac{l}{2}}} (\pi^2 q)^{-\frac{1}{8}} \exp{-\ii \frac{\pi}{16}}\\
    \braket{(2l+1)' | 0'(q)} &\approx (-1)^l \frac{\sqrt{(2l)!}}{l! 4^{\frac{l}{2}}} (\pi^2 q)^{-\frac{1}{8}} \exp{\ii \frac{\pi}{16}} 
\end{align}
Importantly, we find that for all Fourier modes (given that $q$ is large enough so as to be far beyond the relevant exceptional point) the projections of the lowest mode onto them does indeed show the $q^{-1/8}$ scaling found numerically~\cite{kuerstennumerical}, i.e. $\beta_\text{high}=-\frac{1}{8}$. Thus, the high activity limit (and continuum $Q\to\infty$) is indeed expected to be universal with respect to the symmetry of the interaction and not affected by the underlying polar symmetries.

\section{Application to interactions}

The implications for interacting systems are most-straightforward, if we consider the simplest closure, that is consider mean-field dynamics. In this case, the full BBGKY-hierarchy~\cite{balescu,dorfman,ihle2023,boltz2024,boltz2026} is reduced to an equation for the one-particle function.  A simple choice~\cite{peruani2008,boltz2024} for the interactions in eq.~\eqref{eq:model} consists of short-ranged Kuramoto-type couplings, i.e. $\mathcal{F}_{ij} \propto \sin{n(\theta_j-\theta_i)}$ and $\Omega_i=\{j\neq i | \lvert \vec{r}_i-\vec{r}_j\rvert < \ell\}$. The integer parameter $n=1,2,\ldots$ allows to tune the symmetry of the interactions. In ref.~\cite{kuerstennumerical}, Kürsten considered a ring-kinetic closure for a metric-free interaction model, which would by far exceed the scope of this work. We aim to present the core of Kürsten's argument. A similar analysis that explicitly considers the effect of the spatial cut-off has been performed by \citeauthor{escaff2020}~\cite{escaff2020}.

The adapted version of eq.~\eqref{eq:fp3} in mean-field approximation of the metric Kuramoto-like interactions is then
\begin{align}
\partial_t f(\vec r, \phi,t)&=\mathcal{L} f(\vec r,\phi) + J^\text{MF} \\
J^\text{MF} &= \Gamma \partial_\phi \int \sin(n(\theta-\phi)) f(\phi) f(\theta) \,\mathrm{d}\theta
\end{align}
wherein $J^\text{MF}$ is the mean-field collision integral and $\Gamma$ is the dimensionless effective coupling strength. We take the same steps before, that is Fourier transforming with respect to the spatial variables and choosing $\vec k =(k,0)^\mathrm{T}$ as well as Fourier transforming with respect to the double internal angle $\psi=2\phi$. Different to before, we do not take an eigenmode approach to the dynamics, but consider the dynamics of the Fourier modes $\hat{f}_m$. This leads to
\begin{align}
    \begin{split}
\partial_t \hat{f}_m =& -m^2 \hat{f}_m + \frac{\ii q}{2}(\hat{f}_{m-2}+\hat{f}_{m+2}) \\&+ \Gamma m \pi (\delta_{2n}^w - \delta_{-2n}^w) \hat{f}_w \hat{f}_{m-w}
    \end{split}\end{align}
    with Einstein summation over the index $w$.
To limit the analysis here to the case of dynamics that are linear in the modes, we consider perturbations around the disordered state in which only the zeroth mode which is proportional to the density is of order 1. Such a hydrodynamic expansion, leads (after performing the same reduction $f_m\to f_{m'}$ as before, but dropping the primes) \cite{escaff2020,escaff2024,escaff2025,escaff2025b}
\begin{align}
    \begin{split}
\partial_t \hat{f}_m \approx& -m^2 \hat{f}_m + \frac{\ii q}{2}(\hat{f}_{m-1}+\hat{f}_{m+1}) \\&+ \tilde{\Gamma} (\delta_{n}^m   - \delta_{-n}^m) \hat{f}_{m}
\end{split}\end{align} 
with a new effective non-dimensional coupling constant $\tilde{\Gamma}$. Thus, we see as was implied before that the interaction leads to an excitation of the $n$-mode that corresponds to the symmetry of the interaction. Concentrating on the stability of the slowest eigenmode of the free part, we find that the instability is given by
\begin{align}
0 \approx& -\lambda_0 + \tilde{\Gamma}  c_0^{(n)}   
\end{align}
where $c_0^{(n)}$ is the projection of the slow eigenmode onto the $n$-th Fourier mode. In the rest of this work, we consider the case of $n=1$ that is most relevant. This directly gives the following result for the dimensionless coupling strength
\begin{align}
\Gamma_c \sim q^\gamma \sim q^{\alpha-\beta_n} \text{.} \label{eq:gamma}
\end{align}
In our notation, we explicitly highlight that $\beta$ does depend on $n$. For $n=1$ this reproduces the two criticalities found in numerical work~\cite{kuerstennumerical,zhao2025}, whereas this analysis does predict slightly different criticality for higher symmetries. In particular, for small $q$ and non-polar interaction, we predict $\Gamma_c \sim q^{0}$. We summarize the findings as well as the predictions for interacting systems in table~\ref{tab:prediction}.

\begin{table}
    \begin{tabularx}{\linewidth}{|*{3}{Y|}} \toprule
    \multirow{2}{*}{Symmetry} & low activity  & high activity  \\
    & $k \Pe \ll 1$ & $k \Pe \gg 1$\\ \midrule
    polar ($n=1$) & $\begin{array}{cc}
    \\ \alpha = 2 \\ \\
    \beta = 1 \\ \\
    \gamma = 1 \\ ~ 
    \end{array}$& \multirow{2}{*}{$\begin{array}{cc}
    \\ \alpha = \frac{1}{2} \\ \\
    \beta = -\frac{1}{8} \\ \\
    \gamma = \frac{5}{8}  \\ ~
    \end{array}$} \\ \cline{1-2}
    higher ($n>1$)  & $\begin{array}{cc}
    \\ \alpha = 2 \\ \\
    \beta = 2 \\ \\
    \gamma = 0  \\ ~
    \end{array}$ & \\\bottomrule
    \end{tabularx} 
    \caption{ Summary of our findings as they pertain to the scaling of the critical coupling at the onset of instability in interacting systems. The exponents describe the scaling of the slowest mode $\lambda\sim q^\alpha$, see eq.~\eqref{eq:lambda}, the projection of the relevant Fourier mode onto the slowest mode $c_0 \sim q^\beta$, see eq.~\eqref{eq:proj} and the critical coupling $\Gamma_c \sim q^\gamma$, see eq.~\eqref{eq:gamma}, as a function of the effective activity parameter $q\propto k \Pe$. The behavior splits up into two regimes of high (in the continuum limit, $Q\to\infty$) and low activity. In ref.~\cite{kuerstennumerical} the following values were found by 
    numerical diagonalization: $\alpha=2.000000094345308$, $\beta=1.000000008775887$ for low activity and $\alpha=0.5001363406591083$, $\beta=-0.12447191974564421$. Additionally, agent based simulations in ref.~\cite{kuerstennumerical} were found to be consistent with $\gamma=1$ for low activity and $\gamma=5/8$ for high activity. In ref.~\cite{zhao2025}, $\gamma\approx 0$ for low activity and $\gamma \approx \frac{2}{3}$ where determined from agent-based simulations. The deviation between our and Kürsten's result of $\gamma=0.625$ and the value of $\gamma\approx 0.66$ found in this work is not substantial and explained by the resolution of the performed parameter scan. These results are for $n=1$. To our knowledge, there are no existing simulation results for higher interaction symmetry. 
    }
    \label{tab:prediction}
\end{table}

We are not aware of any existing work testing these predictions. In light of the discussion below with respect to mode selection~\cite{escaff2020}, we therefore propose that adapting the model and methodology of \citeauthor{zhao2025}~\cite{zhao2025} to nematic or higher interactions would be a valuable test of our predictions: numerically explore the phase behavior of self-propelled particles subject to diffusional rotation with metric-free interactions that have nematic symmetry over large regions in the parameter space of Peclét number and coupling strength. We plan to explore this in future work. In principle, our results could also tested in experiments. However, having the opportunity to fine-tune activity and coupling over a sufficiently large regime so as to infer scaling will in general be very challenging. Qualitatively identifying substantially different criticialities for small and large activity or, correspondingly, noise could potentially be feasible.

Observing the high-activity limit, requires studying systems at large couplings. In this limit, particle interactions can become effectively irreversible which might lead to large inter-ensemble fluctuations in the transition that could lead to a deviation from our predictions. Practically, one remedy to this could be to also include spatial diffusion into the dynamics in eq.~\eqref{eq:model}. From scaling, one would expect this diffusion to be relevant on lengthscales below the hydrodynamic length scale $\ell_{\text{diff}}\sim \sqrt{D/D_r}$ wherein $D$ is the spatial diffusivity. We consider a coarse-grained mesoscalic description on scales of the interaction length $\ell$, such that most of our methodology is expected to hold as long $\ell \gg \ell_{\text{diff}}$ in the sense that orientational diffusion is the relevant source of noise. A small spatial diffusivity will, however, still suffice to limit the typical interaction times to finite values allowing for reversible coupling of flocking particles. Within our formalism, the spatial diffusion shifts all eigenvalues by $\frac{D}{D_r \ell^2} k^2$. In the high activity limit, this will not affect our predictions of scaling, whereas it does imply a small change in the scaling to $\lambda \sim \frac{D}{D_r \ell^2} k^2$ instead of $\lambda \sim q^2 \sim Pe^2 k^2$ in the low-activity limit as the small shift in the eigenvalues due to the spatial diffusion is relevant for small critical couplings.

\section{Discussion}

With this work, we want to highlight a couple of concepts. Specifically, there is robust analytical foundation to the numerically found scaling behavior~\cite{kuerstennumerical} that underlies the critical scaling relations~\cite{zhao2025} found in metric-free flocking transitions. Analytical access to these exponents, especially in the singularly perturbative high activity limit, is rooted in a mapping of the free Fokker-Planck equation of an active Brownian particle to the imaginary Mathieu equation~\cite{ziener2012}. Our analytical insight allows us also to quantitatively assess the universality of the numerical results. In particular, we find that the high activity limit is indeed similar for all Fourier modes (albeit with varying thresholds as to what qualifies as high), but the case of polar interactions is special in the low activity regime and we, therefore, predict different criticality in this low activity regime for interactions of higher than polar symmetry. 

The existence of simple power-laws has two different origins. In the low activity regime, it is a result from perturbation theory. Treating a small amount of activity as a perturbation around a thermal, equilibrium theory is a standard approach, so this is to be expected. In the high activity regime, however, this is not possible. On a conceptual level, this is evident from the fact that the diffusivity as a small parameter is the coefficient to the highest derivative in the Fokker-Planck equation. Starting from a deterministic active system, an added noise term acts as a singular perturbation. On a practical level, we use the opportunity to show that in the continuum limit the perturbation by the diffusive terms is never small. The simple rational exponents here are rooted in the special topology of the spectrum of the imaginary Mathieu equation, in particular a series of exceptional points. While exceptional points have attracted a lot of interest within active matter research over the last years, this has been limited to studies close to a single exceptional point. In this system, however, there is a robust scaling far from an exceptional point due to coupling to a multitude of exceptional points. To our knowledge, this is the first time this mechanism has been reported as the origin of a critical exponent. We can therefore put exact values to the rare non-equilibrium scaling relations found earlier numerically in refs~\citenum{kuerstennumerical,zhao2025}.

While the application to predict criticality in noisy active aligning systems is very relevant, this is not the only physical result of this work. The mere existence of exceptional points in the relaxational spectrum of the free Fokker-Planck-operator has severe implications as it implies phenomenology akin to a dynamical phase transition~\cite{teza2023} upon variation of the particles' speed or their diffusivity. While there is some work on noise-driven instabilities in interacting self-propelled particles~\cite{escaff2024,escaff2025,escaff2025b}, this aspect of their free dynamics has to our knowledge not received any attention so far.

For a full assessment of the implications of our result, we have to comment on the wave-length dependence. Our results are entirely in terms of an effective activity parameter that is the product of the considered wave-number and the Péclet number. The actual mode selection is subject to particle dynamics and in particular there will in general be a spatial mode-coupling that is not reflected in this work. The alignment interactions will in general have a spatial dependence, for example by introducing a cut-off that separates ``neighboring'' particles from the rest. In systems where this important (density-order coupling, density waves in metric Vicsek models), it will not suffice to simply identify the unstable mode with the system-spanning excitation $k=\frac{2\pi}{L}$. Instead this requires a spatially resolved kinetic theory that accounts for contributions from gradients of the particle density on the scale of the interaction radius. These additional derivatives would then introduce additional wave-number dependencies into the instability analysis. A metric-free interaction as considered by refs.~\cite{kuerstennumerical,zhao2025} should be particularly well-suited for the methodology of this work as these models are designed to suppress the role of density-order coupling. On the other hand, we would expect particularly relevant new terms when considering interactions that directly couple the alignment to the relative displacement of the interaction partners in space~\cite{caussin2014,das2024,kohler2025}. Exploring this and also the role of non-reciprocal interactions, that naturally occur in these systems, and generically lead to traveling states and a rotation of the Fourier modes in the complex plane is a highly interesting route left to future work. As a back-of-the-envelope argument for the change induced by these models, we consider a cohesion/turning-away type interaction~\cite{couzin2002,couzin2005,das2024,shea2025} in which particles align not to (or away from) the orientation direction of their neighbor by to the direction of the displacement vector between them. That is there is angle $\phi_{ij}$ with $\vec{r}_j - \vec{r}_i \propto (\cos \phi_{ij},\sin\phi_{ij})^\mathrm{T}$ and the interaction is of the form $\dot{\phi_i}\propto \sin(\phi_{ij}-\phi_i)$. Due to the symmetry of the problem, there is no mean-field contribution to the kinetic equations in a spatially homogeneous system as contributions from $\phi_{ij}=c$ and $\phi_{ij}=c+\pi$ annihilate each other. We will not explore higher-order contributions~\cite{kohler2025,kohler2026} here, but instead consider a small perturbation around this spatially homogeneous system. A gradient in the density will lead to an effective alignment with respect to the gradient direction within mean-field dynamics and, thus, to an excitation of the ordering that is proportional to the wave-number $k$. Consequently, we would expect that the scaling of the critical coupling strength can no longer be understood in terms of one parameter $q$, but the Peclét number and the wave-number enter with different powers. In particular, we find $\Gamma_c \sim \Pe^{\alpha-\beta} k^{\alpha-\beta-1}$. Interestingly, this would mean that the critical coupling strength is flat in the wave-number (scale-invariant, similar to a type III instability in the Cross-Hohenberg nomenclature~\cite{cross}) in the low activitiy regime and even decreasing (short-wave-length instability similar to the Kelvin-Helmholtz instability~\cite{tong}; importantly this description has a consistency cutoff $k\lessapprox 2\pi$ in our non-dimensional units). While this is certainly intriguing, more detailed considerations in future work are definitely warranted. In particular, we have not considered the role of spatial forces here which are definitely relevant in many experimental realizations~\cite{das2024,fersula2026}, while similar phenomenology can also be achieved with purely orientational interactions (torques)~\cite{kohler2025}. 

The issue of spatial forces is even more pertinent in a different class of variations that has gained considerable attention over the last years: self-(anti-)alignment~\cite{baconnier2022,benzion,musacchio2025,baconnier_2025,ketzetzi2025, tang}. In our dynamics of eq.~\eqref{eq:model}, we considered overdamped dynamics in which the velocity is directly affected by the self-propulsion. However, in many systems the velocity is the result of external forces (hard-core interactions, friction) and the internal self-propulsion whose direction (for example by means of non-isotropic friction) tends to (anti-)align with the velocity direction. Here, one would consider dynamics of the form $\partial_t \vec{r}_i = \vec{v}_i$, $\partial_t \phi_i = \sqrt{2 D_r} \xi_i + \eta \sin(\psi_i-\phi) $ and an model-dependent additional equation for the actual velocity $\vec{v}_i \propto (\cos\psi_i,\sin\psi_i)^{\mathrm{T}}$. While a mean-field equation, for example by way of considering a Dean equation with additive noise~\cite{dean1996,illien2025}, may be accessible and there are ways of incorporating additional knowledge such as the static pair correlation function~\cite{soto} into the kinetic theory formalism to account for hard-core interactions, a full treatment would be outside the scope of this work. However, we expect the overall effect to be similar to what we laid out for the cohesion type interaction. A rather successful effective theory approach for systems with hard spatial forces, is the notion of a density-dependent velocity~\cite{schnitzer1993,tailleur2008,fily2012,bialke_2013,cates2015,Cates_2025} on a coarse-grained scale. This alone can lead to new phenomenology such as motility-induced phase separation. If we, however, assume that this is not relevant and the instability of interest is indeed of a flocking type, the effect is similar to the cohesion type force in the sense that there will be an (anti-)alignment in the direction of density gradients. This leads us to (model details not-withstanding) conclude that the adapted predictions of the preceeding paragraph could hold in these cases.

More generally, self-alignment is an example of a higher-dimensional theory with additional degrees of freedom, that when projected~\cite{zwanzig} to the two degrees of freedom discussed in eq.~\eqref{eq:model} would lead to the emergence of memory-kernels. This is for example also true for inertial~\cite{attanasi2014,cavagna2015} dynamics, models with sensory delay~\cite{loos2019,piwowarczyk_2019,holubec2021} or dynamics with colored noise~\cite{caprini2022} via Gaussian embedding ~\cite{bao2005,siegle2010,siegle2011}. While effective theories in the spirit of our discussion might be efficient, a full theory will always involve dynamical mode-coupling which our treatment here does not account for. 

We do want to reiterate that the interaction sets a length-scale and, therefore, is responsible for a dynamical mode selection. Therefore, our results do correspond to a transition at a finite coupling strength even in the limit of large systems. The derived instability criterion does indeed take the form of a long-ranged, hydrodynamic instability (similar to the Simha-Ramaswamy instability~\cite{simha}) with the coupling strength vanishing with the considered wave-number, but not all modes are explored dynamically.

The results here cover a predominant share of the active models in the literature as it considers self-propelled particles in two dimensions with constant speed. We expect that the bulk of our results applies for models with varying speeds as long as the effective speed distribution is narrow and has sharply decreasing tails. This is due to the fact, that we are considering behavior far from the exceptional points instead of very close, so we expect a certain amount of robustness to model variations. In models where the speed is directional dependent (for example to incorporate the role of particle orientation in acoustic propulsion~\cite{voss}), the free Fokker-Planck-operator will differ, but at the heart of the issue will still be a differential equation of Hill type~\cite{hill1877,magnus2013,dlmf}. This means that the overall mathematical structure should be similar enough that exact asymptotics results similar to those we used here could be achievable.

As another outlook for future work, we note the fact that the brief section on interacting systems only considers mean-field dynamics. While higher order kinetic theory would be available~\cite{kuersten2021,ihle2023,boltz2024}, this is sufficient to get to the core of Kürsten's argument~\cite{kuerstennumerical} and to find the scaling relations for the critical coupling strength. This is important, because it suggests that these exponents could have been determined by the direct numerical solution of kinetic equations. In many situations, the mean-field equations are readily available and their numerical integration with full spatial dependence is feasible~\cite{ihle2013,thueroff,mihatsch2024}. As discussed before, one finds for certain models that the collision integral vanishes to mean-field order~\cite{das2024,kohler2025,kohler2026} for symmetry reasons. In this case, kinetic equations beyond mean-field would definitely be needed to be considered. Identifying the parameter dependence of instabilities in this approach could be a valuable addition to the resource-heavy direct approach of performing parameter scans in long time simulations of large agent based systems. While this is very much in the spirit of field-theoretic approaches like dynamical density functional theory, we are not aware of any works along this direction based on mean-field (or higher) kinetic equations that are derived from microscopic models.

\begin{acknowledgments} 

We thank R.~Kürsten for an inspiring talk about his work~\cite{kuerstennumerical} at the University of Greifswald in 2025 that started this project. We thank Y.~Zhao for a valuable question about spatial diffusion.\nocite{moir2021}
\end{acknowledgments}

\bibliography{lit.bib}
\end{document}